\documentclass[aps,pra,showpacs,twoside,twocolumn,10pt]{revtex4-1}
\usepackage[colorlinks=true, citecolor=red, urlcolor=blue ]{hyperref}
\usepackage{epsfig,newlfont,amssymb,amsfonts,amsmath,bm,subfigure,palatino,mathtools,amsthm,braket,soul,enumitem,color}
\usepackage[normalem]{ulem}

\newcommand{\stkout}[1]{\ifmmode\text{\sout{\ensuremath{#1}}}\else\sout{#1}\fi}

\begin{document}
\title{
Isolating noise and amplifying signal with quantum Cheshire cat}
\author{Ahana Ghoshal$^1$, Soham Sau$^{1,2,3}$, Debmalya Das$^{4,5}$, Ujjwal Sen$^1$}
\affiliation
{$^1$Harish-Chandra Research Institute,  A CI of Homi Bhabha National Institute, Chhatnag Road, Jhunsi, Prayagraj 211 019, India\\
$^2$Department of Physics, School of Physical Sciences, Central University of Rajasthan, Bandarsindri, Rajasthan, 305 817, India\\
$^3$RCQI, Institute of Physics, Slovak Academy of Sciences, Dúbravská Cesta 9, Bratislava 84511, Slovakia \\
$^4$Department of Physics and Astronomy, University of Rochester, Rochester, New York 14627, USA\\
$^5$Center for Coherence and Quantum Optics, University of Rochester, Rochester, New York 14627, USA}

\begin{abstract}
The so-called quantum Cheshire cat is a phenomenon in which an object, identified with a ``cat'', is dissociated from a  property of the object, identified with the ``grin'' of the cat.
%
We propose a thought experiment, similar to this phenomenon, 
with an interferometric setup,  where a property (a component of polarization) of an object (photon) can be separated from the object itself and can simultaneously be amplified when it is already decoupled from its object. We further show that this setup can be used to dissociate two complementary properties, e.g., two orthogonal components of polarization of a photon and identified with the grin and the snarl of a cat, from each other and one of them can be amplified while being detached from the other. Moreover, we extend the work to a noisy scenario, effected by a spin-orbit-coupling --like additional interaction term in the Hamiltonian for the measurement process, with the object in this scenario being identified with a so-called confused Cheshire cat. We devise a gedanken experiment in which such a ``confusion'' can be successfully dissociated from the system, and we find that the dissociation helps in the amplification of signals. 
\end{abstract}

\maketitle

\section{Introduction}
In recent times, a technique known as weak-value amplification has been 
used to amplify 
weak signals~\cite{U}. The method relies on extracting information from a system while minimally disturbing it~\cite{Barchielli1982, Busch1984, Caves1986}.  This weak measurement~\cite{AAV, Duck} of the system is performed using a weak coupling strength between the system and a meter.
 In the weak-value amplification method, a quantum system is initially prepared in a pure state, known as the pre-selected state, following which an observable is weakly measured. After the weak measurement of the observable, a strong measurement of a second observable is carried out on the principal system and a quantity called weak value~\cite{AAV,Duck,YA} is defined by post-selecting an outcome of the second measurement. The weak value of the observable is basically the average shift in the meter readings for the weak measurement corresponding to the post-selected state. Experimental observations of weak values have been reported in Refs.~\cite{Correa, Denkmayr, Sponar, Ashby, Pryde, Hosten, Lundeen, Cormann}.

Although the present work discusses weak values in relation to signal amplification, it may be worthwhile to mention a few other applications of the idea.
Weak values can be used in the direct measurement of a photon wave function~\cite{Lundeen, L2} to measure the spin Hall effect~\cite{Hosten}, in quantum state tomography~\cite{tomo1, tomo2}, in the geometric description of quantum states~\cite{geometry}, and in state visualization~\cite{sv}. It also finds application in quantum thermometry~\cite{qt}, and measuring the expectation value of non-Hermitian operators~\cite{nonherm1, nonherm2}. 
Weak values have been shown to be acquire complex values~\cite{Jozsa} and weak values play important roles in the two-state vector formalism~\cite{twostate}, in the physical understanding of superoscillations~\cite{Super}, and in separating a quantum property from its system~\cite{Aharonov}. Weak measurements have been used to show a double violation of a Bell inequality by a single entangled pair~\cite{npj} and in quantum process tomography where sequential weak measurements are done on incompatible observables~\cite{m}.
A property of weak values that is of special interest to us is that it can lie outside the eigenvalue spectrum of the observable being weakly measured,  and can even be very large~\cite{AAV, Duck}. This aspect is exploited in weak-value amplification.

A significant concern in  experiments is the minimization of noise in the relevant signal and designing measurement techniques that achieve the same. Setting aside logistics and dependencies on other constraints imposed by the particular scenario, a measurement strategy with 
a significantly reduced noise is typically  favored by experimentalists~\cite{SNR1, SNR2, SNR3, SNR4, SNR5, SNR6}. We consider a situation in which noise may be introduced during the process of weak-value amplification if the Hamiltonian coupling the system and the meter 
has extra undesired terms due to the effect of an environmental element. It therefore becomes  necessary to eliminate these terms or suppress them to obtain the amplification of the signal alone. 

An important ingredient of the strategy we discuss in this paper is a phenomenon known as the quantum Cheshire cat. In this gedanken experiment, based on a modified Mach-Zehnder interferometer, a photon can be detected in one arm while its circular polarization can be detected in the other arm, each being absent in the other arm, by measuring the respective weak values~\cite{Aharonov}. Thus, for a particular combination of pre-selected and post-selected states, the photon ``cat'' can be disembodied from its property ``grin'', leading to the name of the effect being chosen after a magical and enigmatic character in the celebrated literary work, \textit{Alice in Wonderland}~\cite{Alice}.
This counter intuitive phenomenon has been experimentally observed using neutron interferometry in Ref.~\cite{Denkmayr} and using photon interferometry in Refs.~\cite{Correa, Sponar, Ashby, Kim2021}. The concept has been further expanded upon in Refs.~\cite{Bancal, At, Duprey, dynamic}, and different kinds of manipulations of the photon polarization, independent of the photon, have been achieved in Refs.~\cite{DasPati, Liu2020, DasSen, wavepar}.

In this paper, we present a gedanken experiment to amplify a property of a photon at a location, independent of the photon being present at the same location. We also show that the same gedanken setup can amplify one property of a photon independent of another property of the same, where the two properties are complementary and correspond to  non-commuting observables.
We believe this is a technologically and fundamentally useful  amalgamation of the two areas, viz., the quantum Cheshire cat and weak-value amplification.
As a demonstration, we consider a scenario where the polarization degree of the photon interacts with another degree of freedom and results in an additional term in the Hamiltonian that governs the measurement. We call the additional term 
the
spin-orbit coupling of the photon (cat), borrowing the nomenclature from the spin-orbit interaction in the relativistic treatment of an electron's dynamics. This additional term behaves as a noise component that changes the shift of the meter. The meter now gives a deflection proportional to the weak value of an effective observable that is different from the weak value of a polarization component, which was to be measured. To avoid the unwanted disturbances caused by the effect of noise in some amplification techniques, we then formulate a gedanken experiment to separate the required observable from the noisy part. We propose that, using this experimental setup, it is possible to reduce the noise effect on average (as a weak value primarily is) and amplify some quantum property by a weak-value measurement with a certain accuracy. 

\par
The rest of the paper is organized as follows. The ideas of weak measurement and quantum Cheshire cat are briefly recollected  in Sec.~\ref{section:2}. In Sec.~\ref{section:3}, we present the thought experiment, based on the concept of the quantum Cheshire cat, to amplify a property of a photon ($z$-component of polarization) independent of the photon itself. We extend the idea 
to amplify one property ($z$-component of polarization) independently of the other ($x$-component of polarization) for a noiseless ideal situation. 
 In Sec.~\ref{section:4}, we evaluate the weak value of the effective observable, resulting from a noisy scenario, where the weak value is obtained from the shift of the meter, weakly coupled with the system, and where the noise is incorporated as a spin-orbit coupling. We also propose an experimental setup useful to amplify the weak value of the effective observable in that section. 
 We summarize in Sec.~\ref{section: conclusion}. 

\section{Review of weak measurement and quantum Cheshire cat}
\label{section:2}
The weak-value scheme~\cite{AAV,Duck} started with a thought experiment for measuring a spin component of a quantum spin-$\frac{1}{2}$ particle, obtaining a result which was far beyond the range of usual values. In this work, the interaction Hamiltonian is usually taken as
\begin{equation}
\label{eqn:ho}
    H_0=-g(t)\hat{A}\otimes \hat{q},
\end{equation}
where $\hat{q}$ is a canonical variable of the meter that is conjugate to momentum $\hat{p}$, $g(t)$ is a time-dependent coupling function with a compact support near the time of  measurement (normalized such that its time integral is unity), and $\hat{A}$ is the 
observable to be measured. In~\cite{AAV,Duck}, $\hat{q}$ has a continuous spectrum. However, we will consider in this paper, instances where $\hat{q}$ can also have a discrete spectrum.
The weak value of $\hat{A}$ is defined as
\begin{equation}
    A_w = \frac{\Braket{\Psi_f|\hat{A}|\Psi_{in}}}{{\Braket{\Psi_f|\Psi_{in}}}}. 
\label{eq:weak}
\end{equation}

The concept of the quantum Cheshire cat~\cite{Aharonov} is based on weak measurement of observables. Typically in a quantum Cheshire cat setup, as seen in Fig.~\ref{fig1}(a), a photon having horizontal polarization $\ket{H}$ is fed into a $50:50$ beam-splitter $BS_1$ of a  Mach-Zehnder interferometer, creating the pre-selected state
\begin{equation}
    \ket{\Psi_{in}}=\frac{1}{\sqrt{2}}(i\ket{L}+\ket{R})\ket{H},
    \label{eq:psi_in}
\end{equation}
where $\ket{L}$ and $\ket{R}$ represent the left and right path degrees of freedom, respectively. We will consider the states of circular polarization of the photon, denoted by $\ket{+}$ and $\ket{-}$, and given by
\begin{equation}
\label{eq:6}
\ket{+}=\frac{1}{\sqrt{2}}(\ket{H}+i\ket{V}),\quad\ket{-}=\frac{1}{\sqrt{2}}(\ket{H}-i\ket{V}),
\end{equation}
as the computational basis. In particular, therefore, $\hat{\sigma}_z=\ket{+}\bra{+}-\ket{-}\bra{-}$ and $\hat{\sigma}_x=\ket{+}\bra{-}+\ket{-}\bra{+}$. Here $\ket{V}$ denotes the state of vertical polarization. This convention is in accordance with that adopted in~\cite{Aharonov}.
In the two arms of the interferometer, weak measurements of the location of the photon and that 
of the photon's circular polarization are carried out. The photon and its polarization interact weakly with appropriate meter states, resulting in deflections in the latter. This interaction is of the form defined in Eq.~(\ref{eqn:ho}).  Next, an arrangement of a half waveplate $HWP$, phase shifter $PS$, beam-splitter $BS_2$, polarization beam-splitter $PBS$ and detectors $D_1$, $D_2$ and $D_3$, elaborated in Fig.~\ref{fig1}(a), is used to post-select the state 
\begin{equation}
    \ket{\Psi_f}=\frac{1}{\sqrt{2}}(\ket{L}\ket{H}+\ket{R}\ket{V}).
    \label{eq:psi_f}
\end{equation}
For this particular post-selected state, it can be seen that the distributions of the deflected meter states center around the weak values of the observables being measured in the arms. To trace the location of the photon, the meters, which are inserted in the left and right arms of the interferometer, measure the projectors $\hat{\Pi}_L=\ket{L}\bra{L}$ and $\hat{\Pi}_R=\ket{R}\bra{R}$, respectively, and similarly the polarization detectors measure the observables $\hat{\sigma}_z^{L}=\hat{\Pi}_L \otimes \hat{\sigma}_z$ and $\hat{\sigma}_z^{R}=\hat{\Pi}_R \otimes \hat{\sigma}_z$.
The corresponding weak values are
\begin{eqnarray}
     (\hat{\Pi}_L)_w=1, && \quad (\hat{\Pi}_R)_w=0, \nonumber\\
     (\hat{\sigma}_z^{L})_w=0, && \quad (\hat{\sigma}_z^{R})_w=1.
\end{eqnarray}
This indicates that the photon passed through the left arm but its $z$-component of polarization passed through the right arm.


\section{Amplification of polarization of a photon without the photon}
\label{section:3}
\begin{figure*}
\centering
\hspace{.25cm}%
\includegraphics[height=8cm,width=8cm]{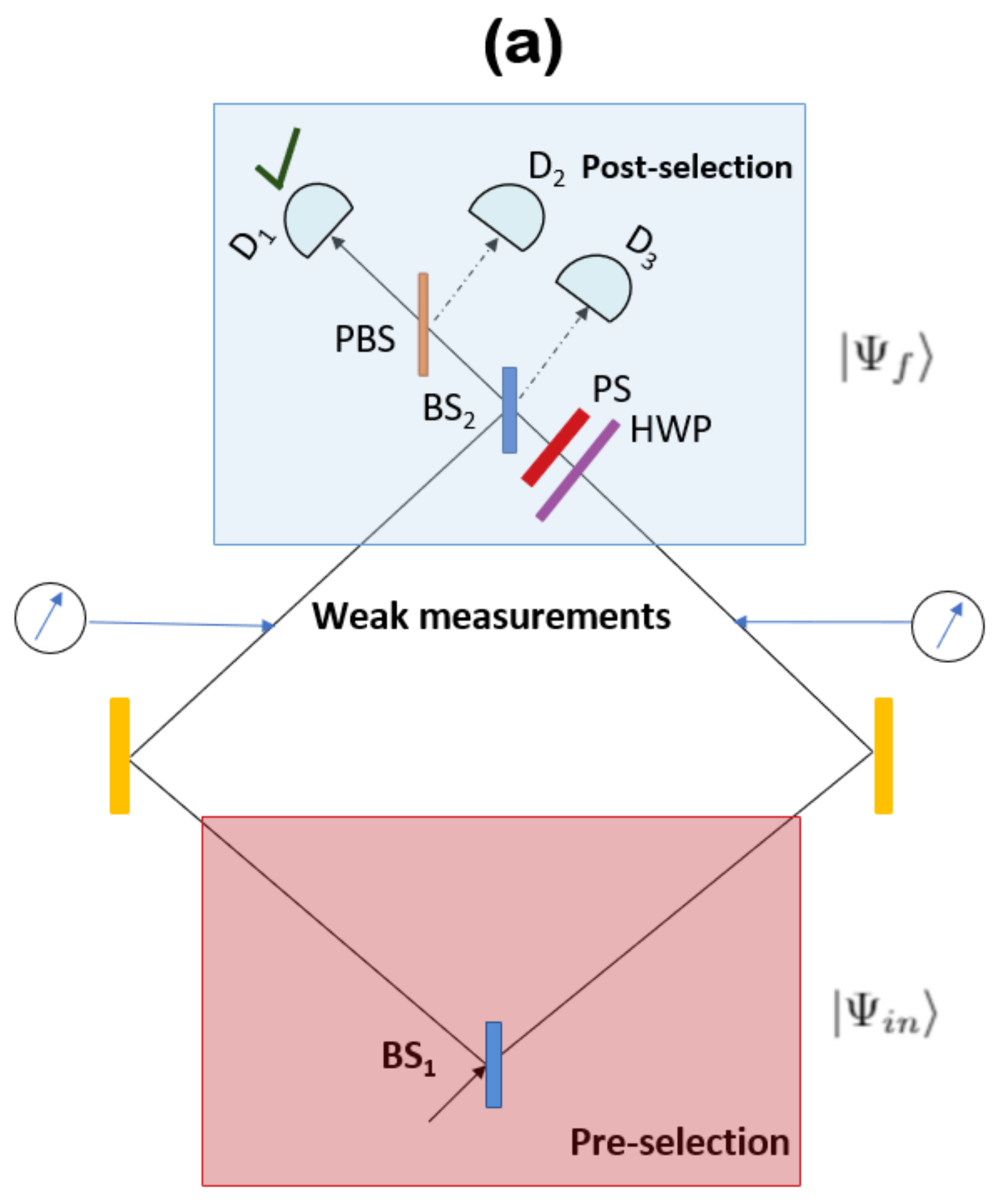}
\includegraphics[height=8cm,width=8cm]{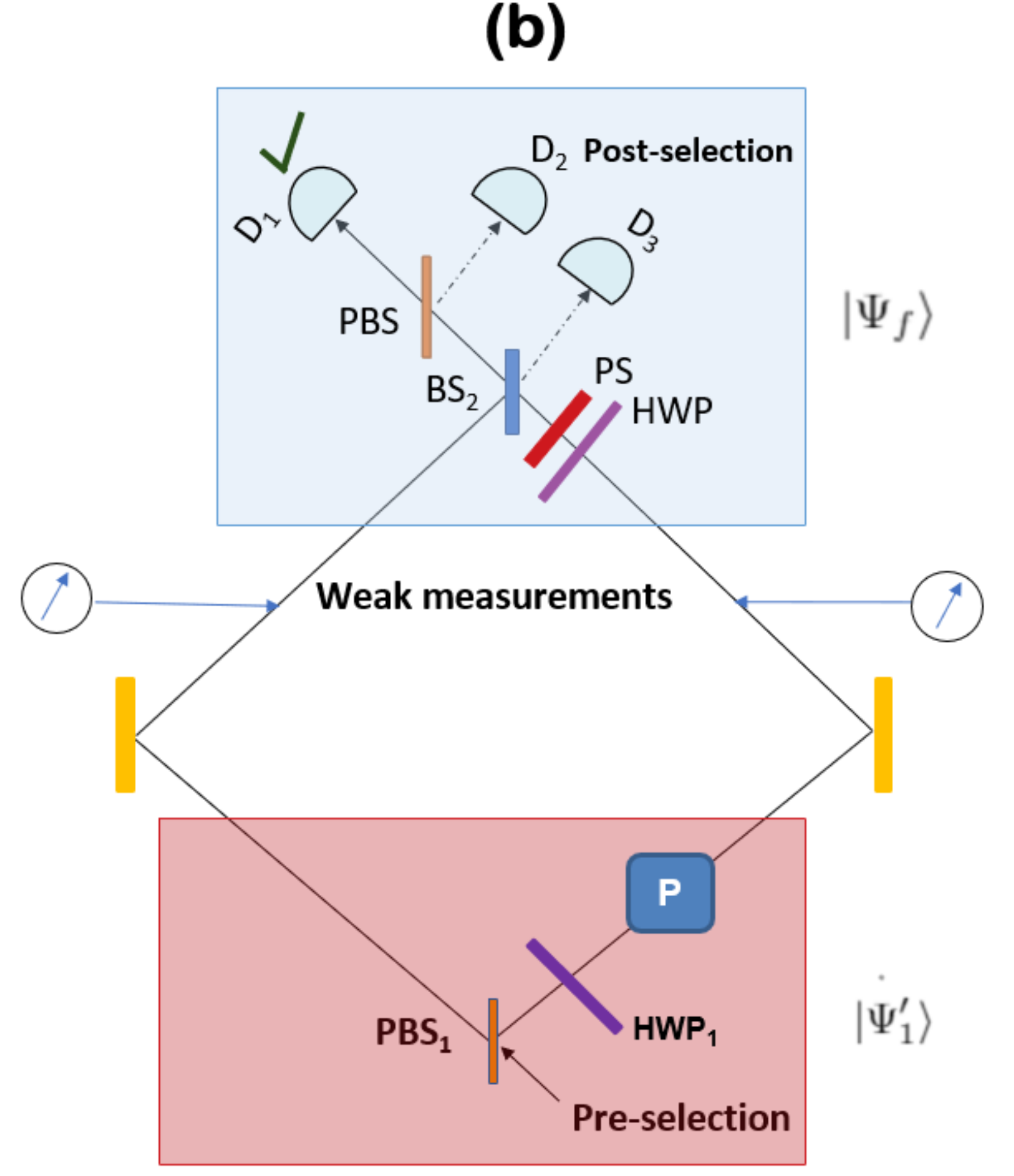}
\caption{Quantum Cheshire cat without and with amplification. (a) Quantum Cheshire cat setup (without amplification). The areas shaded pink and blue carry out the pre-selection and post-selection, respectively. For the latter, only the clicks of detector $D_1$ are selected. Weak measurements of the photon position and the position of polarization are performed by interacting suitable meters weakly, in the two arms of the interferometer. (b) Setup for decoupling the $z$-component of polarization of a photon from the photon itself and amplifying it simultaneously. This configuration is also applicable for dissociating the $z$-component of polarization and the $x$-component of polarization of the photon, and amplifying the former simultaneously.}
\label{fig1}
\end{figure*}

Weak values can be used as a tool for amplifying small signals. Further, we have seen how a property can be separated from a quantum system using the technique of the quantum Cheshire cat. Now we aim to achieve the two phenomena simultaneously, namely, the separation of a property from the object and amplification of the separated property independently of the object. 



The interferometric setup of this gedanken experiment is presented in Fig.~\ref{fig1}(b). Let us begin by considering a photon, propagating along a path degree of freedom state denoted by $\ket{L^\prime}$, in a polarization state $\cos\frac{\theta}{2}\ket{H}+ \sin\frac{\theta}{2}\ket{V}$.
The photon is sent through a polarization beam-splitter $PBS_1$ that transmits horizontally polarized light and reflects the vertically polarized one, leading to the state
\begin{equation}
    \ket{\Psi_1}=\cos\frac{\theta}{2}\ket{L}\ket{H}+ \sin\frac{\theta}{2}\ket{R}\ket{V},
\end{equation}
where $\ket{L}$ and $\ket{R}$ are the two possible photon paths, viz., along the transmitted and reflected paths, respectively, forming the two arms of a Mach-Zehnder interferometer. Note that $\ket{L}$ and $\ket{L^\prime}$ are along the same path after and before the polarization beam-splitter $PBS_1$. In the right arm of the interferometer, we place a half waveplate $HWP_1$ that converts a vertical polarization into a horizontal one, and vice versa, followed by a $\pi$ phase shifter $P$, which introduces a phase $e^{i\pi}$ in the right path. We now have the state
\begin{equation}
    \ket{\Psi^\prime_1}=(\cos\frac{\theta}{2}\ket{L}-i\sin\frac{\theta}{2}\ket{R})\ket{H} . 
    \label{eq: preselect}
\end{equation} 
In the parlance of weak values and the quantum Cheshire cat, this is the pre-selected state.  The post-selection involves a half waveplate $HWP$, a phase shifter $PS$, a beam-splitter $BS$, a second polarization beam-splitter $PBS$, and three detectors $D_1, D_2$ and $D_3$. The working principle is the same as discussed for the quantum Cheshire cat scenario without the amplification requirement. Thus the clicking of the detector $D_1$ can once again be solely selected to obtain the post-selected state
\begin{equation}
    \ket{\Psi_f}=\frac{1}{\sqrt{2}}(\ket{L}\ket{H}+\ket{R}\ket{V}).
    \label{eq:3}
\end{equation}
In the two arms of the interferometric setup, weak measurements of the position of the photon and its circular polarization are performed and the corresponding weak values are obtained. The weak values of the operators $\hat{\Pi}_L$ and $\hat{\Pi}_R$, denoting the positions of the photon in the left and right arms, and $\hat{\sigma}_z^{L}$ and $\hat{\sigma}_z^{R}$, denoting the positions of $z$-components of polarization in the two arms, are then measured to be  
\begin{eqnarray}
(\hat{\Pi}_L)_w&=&\frac{\Braket{\Psi_f|\hat{\Pi}_L|\Psi^\prime_1}}{\Braket{\Psi_f|\Psi^\prime_1}}=1,\nonumber\\
(\hat{\Pi}_R)_w&=&\frac{\Braket{\Psi_f|\hat{\Pi}_R|\Psi^\prime_1}}{\Braket{\Psi_f|\Psi^\prime_1}}=0,\nonumber\\
(\hat{\sigma}_z^{L})_w&=&\frac{\Braket{\Psi_f|\hat{\sigma}_z^{L}|\Psi^\prime_1}}{{\Braket{\Psi_f|\Psi^\prime_1}}}=0,\nonumber\\
(\hat{\sigma}_z^{R})_w&=&\frac{\Braket{\Psi_f|\hat{\sigma}_z^{R}|\Psi^\prime_1}}{\Braket{\Psi_f|\Psi^\prime_1}}=\tan\frac{\theta}{2}.
\label{eq:7}
\end{eqnarray}
Therefore, the photon is detected in the left arm and the $z$-component of polarization is detected in the right arm with a factor which could be amplified by varying the parameter $\theta$. Thus we have achieved the phenomenon of amplifying a property of an object independently of the object: The photon's  polarization component is being amplified in the right arm of the interferometer and the photon is not there.

 This thought experiment can be further extended by separating two orthogonal components of polarization and then amplifying one component independently of the other. Let us consider the two orthogonal components to be the $x$ and $z$-components of polarization, 
 viz. \(\hat{\sigma}_x\) and \(\hat{\sigma}_z\).
The operators $\hat{\sigma}_x^{L}=\hat{\Pi}_L\otimes \hat{\sigma}_x$ and $\hat{\sigma}_x^{R}=\hat{\Pi}_R\otimes \hat{\sigma}_x$ are measured to detect the $x$-component of polarization in the left and right arms of the interferometer, respectively. The corresponding weak values turn out to be
\begin{eqnarray}
(\hat{\sigma}_x^{L})_w&=&\frac{\Braket{\Psi_f|\hat{\sigma}_x^{L}|\Psi_1^{\prime}}}{\Braket{\Psi_f|\Psi_1^{\prime}}}=1, \nonumber\\
(\hat{\sigma}_x^{R})_w&=&\frac{\Braket{\Psi_f|\hat{\sigma}_x^{R}|\Psi_1^{\prime}}}{\Braket{\Psi_f|\Psi_1^{\prime}}}=0.
    \end{eqnarray}
    \par
    When coupled with the weak values of $\hat{\sigma}_z^{L}$ and $\hat{\sigma}_z^{R}$, these indicate that the $z$-component of polarization can be amplified independently of the $x$-component of polarization. 
    The weak value of $\hat{\sigma}_z^{R}$ is seen to acquire a value of $\tan\frac{\theta}{2}$, which means that the weak value will result in outcomes beyond the eigenvalue spectrum in the regions $\theta \in (\pi/2,\pi)$
and $\theta \in (-\pi,-\pi/2)$. 
    
    To realize of the weak-value measurement, 
    the principal system is made to weakly interact with a meter. Let us assume that the meter is initially in a state $\ket{\Phi_{in}}$. Suppose we intend to measure $\hat{\sigma}_z^R$. The weak interaction between the system and the meter 
    can be effected by
     a joint unitary $\hat{U}_{\hat{\sigma}_z^R}=e^{-\int iH_{\hat{\sigma}_z^R}t dt}$ where
     \begin{equation}
         H_{\hat{\sigma}_z^R}=\frac{1}{\sqrt{2}}g(t)[(\hat{I}-\hat{\Pi}_R)\otimes\hat{\sigma}_z\otimes \hat{I} + \hat{\Pi}_R\otimes \hat{\sigma}_z\otimes \hat{q}].
\end{equation}
For the post-selection given by Eq.~(\ref{eq:3}), the meter goes to the state 
    \begin{equation}
    \ket{\Phi_m}=\braket{\Psi_f|\Psi_{1}^{\prime}}[1-ig(\hat{\sigma}_z^R)_w \hat{q}\ket{\Phi_{in}}],
\end{equation}
    with $(\hat{\sigma}_Z^R)_w=1$.
Hence, the meter shows a deflection that is directly proportional to the weak value of the measured observable. 

\section{Amplification of the polarization of a photon in a noisy scenario}
\label{section:4}
In the preceding section, we laid out the procedure for amplifying a property of a quantum system in the absence of the system. We now consider a scenario in which the property we are looking to amplify is affected by noise, in a certain way.
We first present the situation where the weak-value amplification of the noise-affected observable is carried out. As expected, the amplified quantity will contain a contribution from the unwanted noise. We then get rid of the noise by separating it from the signal using a setup similar to that of the quantum Cheshire cat and amplify the signal. 

   
We can conveniently take the observable to be weakly measured, as the $z$-component of polarization $\hat{\sigma}_z$.
We  recall that the measurement of the $z$-component of polarization ideally requires us to set up an interaction of the form
\begin{equation}
H_0=-g(t)\hat{\sigma}_z\otimes \hat{q},
\label{eqn:hoz}
\end{equation}
between the polarization degree of freedom and a convenient meter of our choosing, with meter variable $\hat{q}$. The variable $\hat{q}$ may be a discrete or a continuous meter variable. 
See~\cite{Pryde, npj,m, opticsguo} for experimental realizations of using discrete meter states. 
Let us now consider a  scenario in which there is noise in the interaction Hamiltonian that is analogous to a spin-orbit-coupling term $\hat{L}_x \otimes {\hat\sigma}_x$. The total Hamiltonian is now 
\begin{equation}
    H=-g\delta(t)\hat{I} \otimes \hat{\sigma}_z \otimes  \hat{q}+g^{\prime}  \hat{L}_x \otimes \hat{\sigma}_x \otimes \hat{I}.
\label{eq:Hnoisy}    
\end{equation}
Here $g$ and $g^{\prime}$ are the coupling constants. The coupling between the system and the meter is an instantaneous coupling, while on the contrary, in the second term, there is no dependence of time on spin-orbit coupling. Our aim is to obtain the weak value of $\hat{\sigma}_z$. Due to the noise, the effective observable of the total system, resulting in the deflection in the meter, is different from that in the noiseless case obtained from Eq.~(\ref{eqn:hoz}) in~\cite{AAV,Duck}.

Let us consider the initial state of the system as $\ket{\Psi_{in}}$ and the post-selected state as $\ket{\Psi_f}$. As mentioned before, the meter state could be chosen as a discrete or a continuous spectrum, depending on the type of the experimental setup. In our work, we consider both scenarios: One is with a meter variable $\hat{q}$ considered as discrete and the other is with a continuous state distribution of the same. As an example of a discrete meter state, we take a discrete Gaussian function with the standard deviation $\sqrt{2}\Delta$ as
\begin{equation}
\label{eq:dis}
\ket{\Phi_{in}}_{dis}= \sum_{k=-N}^{N}\exp \Big ( -\frac{q_k^2}{4\Delta^2}\Big)\ket{q_k},
\end{equation}
where we are assuming a discrete meter variable $q_k=k$ with $k$ having the values $0, \pm 1, \pm 2\ldots \pm N$. The corresponding $p$ representation of this meter state turns out to be
\begin{equation}
    \ket{\Phi_{in}}_{dis}=\sum_{l=-N}^{N} \exp \left ( -\Delta^2 p_l^2\right)\xi(p_l)\ket{p_l},
\end{equation}
where $\xi(p_l)=\sum_{k=-N}^{N} e^{-\frac{1}{4\Delta^2}\big(q_k+2i\Delta^2 p_l\big)^2}$ and $p_l=\frac{l}{(2N+1)}$ with $l$ having the values $0, \pm 1, \pm 2\ldots \pm N$.
In the limit of $N \rightarrow \infty$, $\xi(p_l)$ will be independent of $p_l$. 
The parallel example in the continuous case may be taken as 
\begin{equation}
\ket{\Phi_{in}}_{con}= \int_{-\infty}^{+\infty} dq\exp \Big ( -\frac{q^2}{4\Delta^2}\Big)\ket{q},
\label{eq:meter} 
\end{equation}
with the $p$-representation,
\begin{equation}
\ket{\Phi_{in}}_{con}= \int_{-\infty}^{+\infty} dp\exp \Big ( -\Delta^2 p^2\Big)\ket{p}.
\label{eq:meterp} 
\end{equation}
To obtain the $p$ representations from the $q$ representation in both the discrete and  continuous cases, we neglect multiplicative constants.
After post-selection, the final state of the meter is given by
\begin{equation}
  \ket{\Phi_f}=\bra{\Psi_f}\hat{U}(t)\ket{\Psi_{in}}\ket{\Phi_{in}},
\end{equation}
where $t$ is the time of measurement. This is true for both the discrete and continuous meter states and hence we have omitted the subscripts. As the Hamiltonians at different times do not commute, the expansion of $\hat{U}(t)$ requires using the Dyson series expansion, leading to 
\begin{eqnarray}
\label{eq:uni}
    \hat{U}(t)
    &=& 1+ \sum_{n=1}^{\infty} \frac{1}{n!}\Big (\frac{-i}{\hbar}\Big)^n \int_{0}^{t}dt_1 \int_{0}^{t_1}dt_2 \cdots \int_{0}^{t_{n-1}}dt_{n-1}\nonumber\\ &&\phantom{ dure cholo jai}
\times    \mathcal{J}[H(t_1)H(t_2) \cdots H(t_n)].
\end{eqnarray}
We re-define the coupling constants \(g\) and \(g'\) so that we can effectively set 
$\hbar=1$. Using time ordering 
in the expansion,
we have
\begin{eqnarray}
 1&+& (-i) \int_{0}^{t}dt_1 H(t_1)\nonumber\\
&+&\frac{1}{2!}(-i)^2 \int_{0}^{t}dt_1 \int_{0}^{t_1}dt_2 H(t_1)H(t_2)\nonumber\\
 &+&\frac{1}{2!}(-i)^2 \int_{0}^{t}dt_2 \int_{0}^{t_2}dt_1 H(t_2)H(t_1)+ \cdots.
\end{eqnarray}
Interchanging $t_1$ and $t_2$ in the last (displayed) term, we get
\begin{eqnarray}
 1&+& (-i) \int_{0}^{t}dt_1 H(t_1)\nonumber\\
&-& \int_{0}^{t}dt_1 \int_{0}^{t_1}dt_2 H(t_1)H(t_2)+ \cdots.
\label{ramakantakamar}
\end{eqnarray}
Substituting the Hamiltonian $H$ from Eq.~(\ref{eq:Hnoisy}) in this expression, the form of the unitary turns out to be 
{\begin{eqnarray}
\label{eq:approx}
&&1-i({g\hat{I} \otimes \hat{\sigma}_z \otimes \hat{q}+g^\prime t \hat{L}_x\otimes \hat{\sigma}_x} \otimes \hat{I}) -[g^2(\hat{I} \otimes \hat{\sigma}_z \otimes \hat{q})^2 \nonumber\\ 
&+& g g^\prime t (\hat{L}_x \otimes \hat{\sigma}_x\hat{\sigma}_z \otimes \hat{q})+g^{\prime^2} \frac{t^2}{2}(\hat{L}_x \otimes \hat{\sigma}_x \otimes \hat{I})^2]+ \cdots. \quad \quad
\end{eqnarray}}
The measurement time $t$ is chosen from the regime {$\frac{g}{g^{\prime}} \ll t \ll \frac{\sqrt{g}}{g^{\prime}}$}. Thus we can
now assume $g$ and $g^{\prime}$ to be sufficiently small to neglect the $g^2$ term in~(\ref{eq:approx}) and the higher-order terms that are not present in~(\ref{eq:approx}).
On the other hand, we retain the terms containing $gg^{\prime}t$ and $g^{\prime^2}t^2$. Using Eq. 
(\ref{eq:dis}), the final state of the meter after weak measurement and post-selection reads  
\begin{eqnarray}
\label{eqn:zzz}
  \ket{\Phi_f}_{dis} &\approx&  \braket{\Psi_f|\Psi_{in}}\sum_{k=-N}^{N} e^{-\frac{q_k^2}{4\Delta^2}}\Big[1+i g q_k(\hat{\sigma}_z)_w\nonumber\\ 
  &-& i g^{\prime} t (\hat{L}_x\otimes \hat{\sigma}_x)_w-gg^\prime tq_k (\hat{L}_x\otimes \hat{\sigma}_x \hat{\sigma}_z)_w \nonumber\\
  &-& g^{\prime^2}\frac{t^2}{2}(\hat{L}_x\otimes \hat{\sigma}_x)_w^2\Big]\ket{q_k}.
  \end{eqnarray}
The weak values above are obtained using the definition in Eq. (\ref{eq:weak}). Now, in general, for an observable $\hat{A}$  and post-selection of $\ket{\Psi_f}$, with $A_w$ being the weak value of $\hat{A}$, the shifted meter state is given by
\begin{equation}
\label{eq:shift1}
 \ket{\Phi_f}_{dis} \approx \braket{\Psi_f|\Psi_{in}}\sum_{k=-N}^N e^{igq_kA_w}\exp\Big(-\frac{q_k^2}{4\Delta^2}\Big)\ket{q_k}.
\end{equation}
Comparing Eqs. (\ref{eqn:zzz}) and (\ref{eq:shift1}), we get
\begin{eqnarray}
e^{igq_kA_w}&=&1+igq_k(\hat{\sigma}_z)_w-ig^{\prime}t(\hat{L}_x\otimes\hat{\sigma}_z)_w \nonumber\\ 
&&-gg^\prime tq_k(\hat{L}_x\otimes\hat{\sigma}_x\hat{\sigma}_z)_w-g^{\prime^2}\frac{t^2}{2}(\hat{L}_x\otimes\hat{\sigma}_z)_w^2. \quad
\label{shift2}
\end{eqnarray}
 Let
\begin{eqnarray}
a_w&=&-ig^{\prime}t(\hat{L}_x\otimes\hat{\sigma}_z)_w-g^{\prime^2}\frac{t^2}{2}(\hat{L}_x\otimes\hat{\sigma}_z)_w^2\nonumber,\\
A^{\prime}_w&=&(\hat{\sigma}_z)_w+ig^\prime t (\hat{L}_x\otimes\hat{\sigma}_x\hat{\sigma}_z)_w.
\label{eq:weak_noise}
 \end{eqnarray}
So, the final state of the meter can be rewritten as
\begin{equation}
 \ket{\Phi_f}_{dis} \approx \braket{\Psi_f|\Psi_{in}}\sum_{k=-N}^N\; e^{a_w} e^{igq_kA^{\prime}_w}\exp\Big(-\frac{q_k^2}{4\Delta^2}\Big)\ket{q_k},
\end{equation}
with the corresponding $p$-representation being
\begin{eqnarray}
 &&\ket{\Phi_f}_{dis}\approx \braket{\Psi_f|\Psi_{in}}\sum_{l=-N}^N e^{a_w} \exp\Big[-\Delta^2(p_l-gA^{\prime}_w)^2\Big]\nonumber\\
 &&\phantom{ami hridoyer bolite byakul sudha} \times \xi(p_l-gA_w^{\prime})\ket{p_l}\label{eq:30}\\
 &&\approx \braket{\Psi_f|\Psi_{in}}\sum_{l=-N}^N e^{a_w^{\prime}} \exp\Big[-\Delta^2\Big(p_l-gA^{\prime\prime}_w\Big)^2\Big]\ket{p_l}.\label{eq:31}
\end{eqnarray}
Here $a^{\prime}_w$ and $A^{\prime\prime}_w$ are implicitly defined via the expression~(\ref{eq:30}), which is equal to the  expression~(\ref{eq:31}) and $A^{\prime\prime}_w\rightarrow A^{\prime}_w$, $a^{\prime}_w\rightarrow a_w$ as $N\rightarrow \infty$. The factor $e^{a_w^{\prime}}$ does not contribute to the shift of the meter. So the deflection of the meter is proportional to the weak value $A_w^{\prime\prime}$. This $A_w^{\prime\prime}$ is difficult to be given in a closed (explicit) analytic form. In the continuous limit $N\rightarrow \infty$, when the meter state is taken as a continuous one, given in Eq.~(\ref{eq:meter}), the effective observable, measured in the weak measurement conjured by the noisy Hamiltonian in Eq. ({\ref{eq:Hnoisy}}), is given by
\begin{equation}
    A^{\prime}=\hat{\sigma}_z+ig^\prime t \hat{L}_x\otimes\hat{\sigma}_x\hat{\sigma}_z.
    \label{eq:A_w_noisy}
\end{equation}
The steps of the calculation for the continuous meter case are given in Appendix~\ref{discrete}. In the further discussion of our paper we will use the effective observable obtained in the continuous limit. The results for the discrete case will  however be close to those obtained using the effective observable \(A^{\prime}\) in Eq.~(\ref{eq:A_w_noisy}) for large \(N\). Moreover, there are instances below where the discrete and continuous cases match (in form).

If instead of $\hat{\sigma}_x$ the noise term in the Hamiltonian of Eq.~(\ref{eq:Hnoisy}) contains $\hat{\sigma}_z$, as in the cases, 
\begin{eqnarray}
    H_1=-g\delta(t)\hat{I} \otimes \hat{\sigma}_z \otimes  \hat{q}+g^{\prime}  \hat{L}_x \otimes \hat{\sigma}_z \otimes \hat{I},    \nonumber \\
    H_2=-g\delta(t)\hat{I} \otimes \hat{\sigma}_z \otimes  \hat{q}+g^{\prime}  \hat{L}_z \otimes \hat{\sigma}_z \otimes \hat{I},
    \label{eq:noise:parallel}
\end{eqnarray}
then by calculating the effective observable in a similar fashion, we get, respectively,
\begin{eqnarray}
    &&A_1^{\prime}=\hat{\sigma}_z+ig^\prime t \hat{L}_x\otimes\hat{I},\nonumber \\
    &&A_2^{\prime}=\hat{\sigma}_z+ig^\prime t \hat{L}_z\otimes\hat{I}.
\end{eqnarray}

We can also consider a noisy interaction Hamiltonian of a different form. Precisely, we can 
take the noisy part of the interaction Hamiltonian to be a three-body term so that 
it is coupled to the meter with a coupling parameter $g(t)$ which has a compact support near the measurement time $t$: 
    \begin{eqnarray}
        H^\prime=-g(t)\big(\hat{I}\otimes\hat{\sigma}_z \otimes \hat{q}-\hat{L}_x\otimes\hat{\sigma}_x \otimes \hat{q}\big).
        \label{eq:Hnoisysimple}
    \end{eqnarray}
Using the same method as before, we see that the effective observable resulting in the shift of the meter is 
\begin{equation}  
    A^\prime=\hat{\sigma}_z-\hat{L}_x\otimes\hat{\sigma}_x
    \label{eq:A_w_1_noisy}.
   \end{equation}


\begin{figure}
\includegraphics[height=4cm,width=8cm]{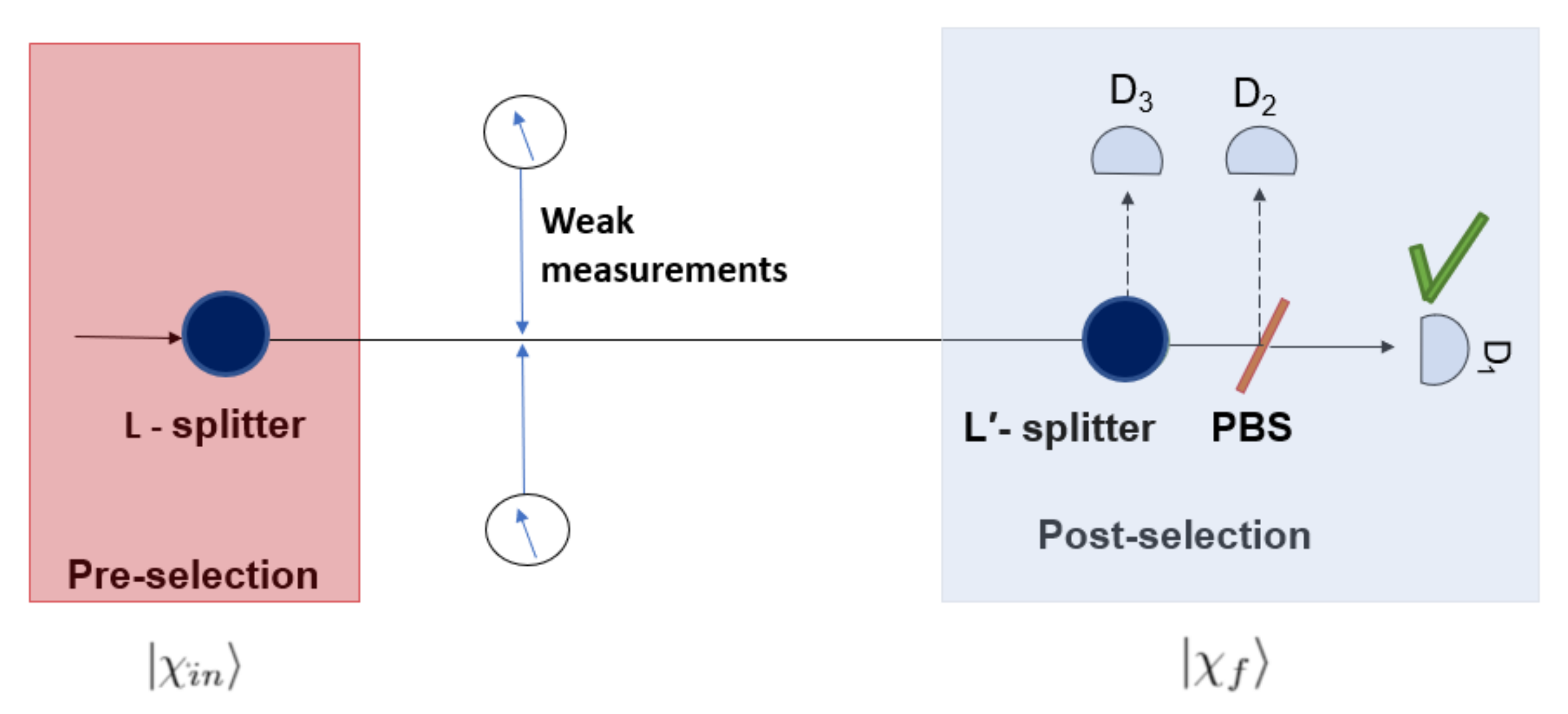}
\caption{Setup for amplifying the weak value of the effective observable in the noisy scenario. The $L$-splitter adds an orbital angular momentum degree of freedom to the physical system. The $L^\prime$-splitter transmits the component of orbital angular momentum parallel to $\ket{v_a}$ and reflects any component orthogonal to it.}
\label{fig2}
\end{figure}

To demonstrate 
the working principle in either case [i.e., when the interaction Hamiltonian is given by either Eq.~(\ref{eq:Hnoisysimple}) or~(\ref{eq:Hnoisy})],
let us consider a set of pre-selected and post-selected states as follows:
\begin{eqnarray}
\ket{\chi_{in}}&=&\frac{1}{\sqrt{2}}(\ket{v_a}+i\ket{v_b})\otimes \ket{H},\nonumber\\
\ket{\chi_f}&=&\ket{v_a} \otimes(\cos\alpha\ket{H}+\sin\alpha\ket{V})
\label{pre_post_noise_1}.
    \end{eqnarray}
In these states, the first degree of freedom represents the angular momentum component \(\hat{L}_x\), while the second degree of freedom represents polarization. We assume that the orbital quantum number \(l\) is conserved, for the system under study, at 1. Correspondingly, the dimension of the space spanned by the eigenvectors of \(\hat{L}_x\) is three dimensional.
For simplicity,
we work in a scenario where one of the dimensions in this three-dimensional space is naturally or artificially forbidden. 
The remaining 
two-dimensional Hilbert space is spanned by orthonormal 
vectors $\ket{v_a}$ and $\ket{v_b}$, with $\hat{L}_x$ represented as
\begin{equation}
    \hat{L}_x=-i(\ket{v_a}\bra{v_b}-\ket{v_b}\bra{v_a}),
\end{equation}
where 
    \begin{eqnarray}
        \ket{v_a}= \frac{1}{\sqrt{2}}\begin{pmatrix} 1 \\ 0 \\ 1
        \end{pmatrix}  , \quad
        \ket{v_b}= \begin{pmatrix} 0 \\- i \\ 0
        \end{pmatrix}, 
    \end{eqnarray}
expressed in the eigenbasis of \(\hat{L}_z\). To prepare the pre-selected state, we can send a photon with an initial polarization $\ket{H}$ through an arrangement, which we call the $L$-splitter  in Fig.~\ref{fig2}, where it acquires an angular momentum of $\frac{1}{\sqrt{2}}(\ket{v_a}+i\ket{v_b})$. We intend to measure the weak value of $\hat{\sigma}_z$ and so bring in a meter in the initial state given by Eq.~(\ref{eq:dis}) for the discrete pointer state and Eq.~(\ref{eq:meter}) for the continuous pointer state and set up an interaction of the form in Eq.~(\ref{eqn:hoz}). However, unknown to us, the form of interaction is actually as in Eq.~(\ref{eq:Hnoisy}). 
To get the post-selected state, the photon passes through another arrangement, the $L^\prime$-splitter, that transmits orbital angular momentum $\ket{v_a}$ and reflects any orthogonal component towards the detector $D_3$. The transmitted photon is then passed through a polarization beam-splitter $PBS$, that is chosen so that it transmits light of polarization $\cos\alpha\ket{H}+\sin\alpha\ket{V}$ towards a detector $D_1$ and reflects any orthogonal polarization en route to detector $D_2$. Thus by selecting the clicks of $D_1$ alone, we can post-select the state $\ket{\chi_f}$. 
     
However, due to the inherent noise in the Hamiltonian of the form~(\ref{eq:Hnoisy}), as a result of the extra spin-orbital interaction, the weak value of the effective observable 
$A^{\prime}=\hat{\sigma}_z+ig^\prime t \hat{L}_x\otimes\hat{\sigma}_x\hat{\sigma}_z$ is measured (instead of \(\hat{\sigma}_z\)) to be 
\begin{eqnarray}
 A^\prime_w={(g^\prime t+i)\tan\alpha}.
\label{A_wk}
\end{eqnarray}
In contrast, if the unknown noise is a three-body interaction as in Eq.~(\ref{eq:Hnoisysimple}), the weak value of the effective observable given in Eq.~(\ref{eq:A_w_1_noisy}) turns out to be
\begin{equation}
   A_w^{\prime}=1+i \tan\alpha
    \label{A_wk1}
\end{equation} 
using the setup schematically demonstrated in Fig.~\ref{fig2}.

So, in both noisy situations [Eqs. (\ref{eq:Hnoisy}) and (\ref{eq:Hnoisysimple})] we can amplify the weak value of the respective effective observables [see Eqs. (\ref{eq:A_w_noisy}) and (\ref{eq:A_w_1_noisy})] by varying the parameter $\alpha$ using different $PBS$s. However, due to the presence of spin-orbit coupling in the system, we are unable to determine the weak value of $\hat{\sigma}_z$ as, 
instead of $\hat{\sigma}_z$, the deflection of the meter is proportional to the weak value of an observable which contains an unwanted noise along with $\hat{\sigma}_z$. Hence, while attempting to amplify the weak value of the $z$-component of polarization,
we unintentionally amplify the weak value of a constituent noise. This, e.g., may cause disadvantages in  applications of quantum technologies, which depend on the weak-value enhancement of the polarization degree of freedom of the system. It is plausible that the presence of the noise effects considered is more probable than the ideal noiseless situation, and therefore it will be beneficial to design an experimental setup that can ``disembody'' the noise from the required observable.

\subsection{Disembodiment of noise from the ideal system using quantum Cheshire cats 
}
\label{section:5}

\begin{figure}
\includegraphics[height=8cm,width=8cm]{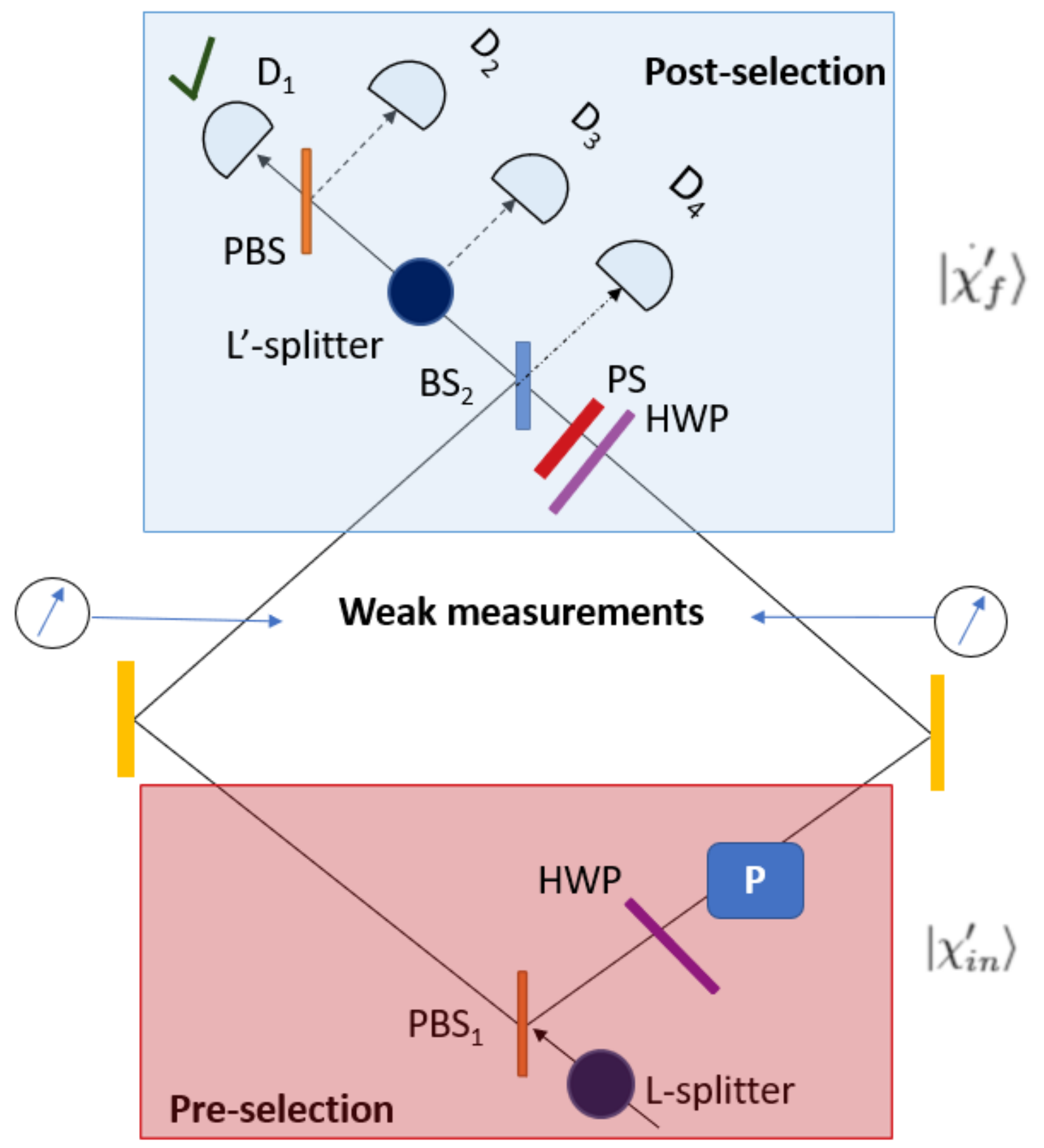}
\caption{
Interferometric setup for separating the spin-orbit-coupling --like noise and simultaneously amplifying the signal corresponding to the chosen observable. See the text for details.
}
\label{fig3}
\end{figure}    

  With the general working principle established earlier in Sec.~\ref{section:2}, we now proceed to get 
  the amplified signal
  of the $z$-component of polarization by disassociating, with the help of the Cheshire cat mechanism. 
   
  The noise originates from the interaction with the unintended degree of freedom (in this case, $\hat{L}_x$) during the weak measurement process. 
  The intended pre-selected and post-selected states are $\ket{\Psi_1^\prime}$ and $\ket{\Psi_f}$, respectively [see Eqs.~(\ref{eq: preselect}) and~(\ref{eq:3})]. However, the photon passing through the $L$-splitter picks up a new degree of freedom, an angular momentum component given by $\frac{1}{\sqrt{2}}(\ket{v_a}+i\ket{v_b})$, as shown in Fig.~\ref{fig3}. So the effective pre-selected state is
    \begin{equation}
       \ket{\chi_{in}^{\prime}}=(\cos\frac{\theta}{2}\ket{L}- 
       i\sin\frac{\theta}{2}\ket{R}\otimes\frac{1}{\sqrt{2}}(\ket{v_a}+i\ket{v_b})\otimes \ket{H}.
       \label{eq:pree}
    \end{equation} 
To achieve the desired outcome, we are required to carry out post-selection in     
\begin{equation}
\ket{\chi_f^{\prime}}=(\cos\alpha\ket{L}\ket{H}+ \sin\alpha\ket{R}\ket{V})\otimes\ket{v_a}.
\label{eq:post}
\end{equation}
To obtain this state, the beam-splitter $BS_2$ needs to be of a transmission coefficient $\cos^2 \alpha$ and reflection coefficient $\sin^2 \alpha$ rather than being the $50:50$ type. Also, we assume that the state passes through a device, the $L^\prime$-splitter, that permits only the orbital angular momentum component along $\ket{v_a}$ towards the $PBS$ and reflects any orthogonal component towards the detector $D_3$, as shown in Fig.~\ref{fig3}.        
        The measured weak values are
        \begin{eqnarray}
        &&(\hat{\sigma}_z^L)_w = 0,\nonumber \\
        &&(\hat{\sigma}_z^R)_w = \tan\frac{\theta}{2}\tan\alpha, \nonumber \\
        &&(\hat{L}_x\otimes\hat{\sigma}_x)^L_w = 1, \nonumber \\
        &&(\hat{L}_x\otimes\hat{\sigma}_x)^R_w = 0. 
        \label{wk_final}
        \end{eqnarray}
Therefore, we can conclude that one can amplify the weak value of the observable $\hat{\sigma}_z$ separating it from the noise part $\hat{L}_x \otimes \hat{\sigma}_x$. Moreover, the enhancement of the weak value of $\hat{\sigma}_z^R$ can be more than that of the effective observables obtained in the two previous cases [described in Eqs.~(\ref{A_wk}) and~(\ref{A_wk1}), using a linear setup, as 
depicted
in Fig.~\ref{fig2}], because 
of the 
extra tuning parameter $\theta$. Therefore, even if we fix the parameter $\alpha$, we are still able to amplify the weak value of $\hat{\sigma}_z$ by changing the parameter $\theta$ with the use of different $PBS_1$s. So now we have successfully achieved our goal of amplifying the weak value of the required observable by splitting up the noise from the system. In addition, the enhancement of the weak value of the required observable is
greater than that of the effective observable detectable in the noisy situation [compare Eqs.~(\ref{A_wk}) and~(\ref{A_wk1}) with Eq.~(\ref{wk_final})]. 
We need to choose 
an initial state of suitable polarization in the preparation of the pre-selected state. Also, to measure the weak values in an experimental procedure, we have to construct a unitary parallel to the one mentioned in the noiseless situation. 
The composite setup of the system and meter can be acted on by the Hamiltonian $H_{\hat{\sigma}_z^R}^{\prime}$ for measuring $\hat{\sigma}_z^R$, where
\begin{eqnarray}
H_{\hat{\sigma}_z^R}^{\prime} &=& g\delta(t-t^{\prime})[(\hat{I}-\hat{\Pi}_R) \otimes \hat{I} \otimes \hat{\sigma}_z \otimes \hat{I}\nonumber\\
&&\phantom{na go ei je dhula}+\hat{\Pi}_R \otimes \hat{I}\otimes\hat{\sigma}_z \otimes \hat{q}]. \label{eq:H_sigma}
\end{eqnarray} 
So the unitary generated by the Hamiltonian $H_{\hat{\sigma}_z^R}^{\prime}$ will be $U_{\hat{\sigma}_z^R}^{\prime}=e^{-\int iH_{\hat{\sigma}_z^R}^{\prime}t dt}$. After the post selection, the meter state turns out to be
\begin{equation}
\ket{\Phi_m}=\braket{\chi_f^{\prime}|\chi_{in}^{\prime}}[1-ig(\hat{\sigma}_z^R)_w \hat{q}\ket{\Phi_{in}}].
\end{equation}
Hence, the deflection of the meter state is proportional to the weak value of $\hat{\sigma}_z$ on the right arm, up to the first-order term in the expansion of the unitary.
The constructed unitary is almost the same as in the noiseless scenario,
the only difference being that we have to incorporate an identity in the degree of freedom of orbital angular momentum. 
Similarly, to measure $(\hat{L}_x\otimes \hat{\sigma}_x)^L$, the unitary is $\hat{U}_{(\hat{L}_x\otimes \hat{\sigma}_x)^L}^{\prime}$, where
\begin{eqnarray}
     H_{(\hat{L}_x\otimes \hat{\sigma}_x)^L}^{\prime} &=& g^{\prime}[(\hat{I}-\hat{\Pi}_L) \otimes \hat{L}_x \otimes \hat{\sigma}_x \otimes \hat{I}\nonumber\\
     &&\phantom{jodi tare}+ \hat{\Pi}_L \otimes \hat{L}_x\otimes\hat{\sigma}_x \otimes \hat{q}],
     \label{eq:H_last}
\end{eqnarray}
and here the final meter state ends up being
\begin{equation}
    \ket{\Phi_m}=\braket{\chi_f^{\prime}|\chi_{in}^{\prime}}[1-ig^{\prime}t(\hat{L}_x\otimes\hat{\sigma}_x)^R)_w \hat{q}\ket{\Phi_{in}}].
\end{equation}
Note that the subscript dis or con has been omitted here, as the same form of the equation is true in both cases.
\par

In practice, a more complex scenario can arise when the noise couples with the measured observable, as in Eq.~(\ref{eq:noise:parallel}). In this case, one may not be able to find a suitable pre- and post-selection to decouple the noise from $\hat{\sigma}_z$. It could be possible to separate $\hat{\sigma}_z$ and $\hat{L}_x \otimes \hat{\sigma}_z$ or $\hat{L}_z \otimes \hat{\sigma}_z$, but in both arms of the interferometer, there still remains a contribution of $\hat{\sigma}_z$. Hence the complete dissociation of the $z$-component of polarization of the photon 
from the noise part may not be achievable by this method.

In the noise model proposed here, the angular momentum degree of freedom of the system is acting as a source of noise and it couples to another degree of freedom of the system, which we want to measure. So it is like a subsystem interacting with the original system. In a realistic noisy scenario, to introduce a generic noise in the system, we take an auxiliary system from outside, operate a global unitary on the composite system-auxiliary setup, and then trace out the auxiliary. The same method can be followed with a different degree of freedom of the system instead of using the auxiliary system. The mathematical modeling will be the same in both instances. There are previous works where the noise is generated from a degree of freedom of the system itself. See, e.g., \cite{Zukowski1999,Zukowski2023}. Also, in \cite{Flores} it was shown that decoherence effects can be generated due to the coupling of mesoscopic variables of the system and internal degrees of freedom of the same. The center of mass of a system can have different degrees of freedom, which may interfere, and such interference can get effectively decohered due to the coupling of the center of mass of the system with the internal vibrational degrees of freedom, as was studied in~\cite{Brun}. See also~\cite{Hillery,Nikolic} in this regard. 

An additional point to be noted here is that  in practical scenarios, the noise source is usually unknown and we have to trace out the subsystem. In our case, for the measurement, we have to concentrate on the meter state, which is done by removing both degrees of freedom of the system from the total system including the meter. This tracing out is performed implicitly in the method of weak measurement. This implicit tracing out of the system has  also been performed in previous papers in this direction, e.g., in~\cite{AAV,Duck}. Note that for this implicit tracing out of the system, the noise source is also being traced out. So no additional explicit tracing out,  corresponding to the noise in the system, is required.
\section{Conclusion}
\label{section: conclusion}
To summarize, we have proposed a thought experiment related to the so-called quantum Cheshire cat in which a component of polarization could be amplified independently of the photon, using interferometric arrangements. 
Furthermore, we extended the setup to a scenario in which, of two complementary polarization components of a photon, one  can be amplified while being detached from the other.
Moreover, we considered a noisy scenario in which the noise is generated by a spin-orbit-coupling --like interaction term in the Hamiltonian governing the measurement process. 
We analyzed the amplification of a chosen observable in the presence of noise, both with and without the noise term being 
dissociated, on average, from the object by using a quantum Cheshire cat--inspired setup. 



It has been pointed out in the literature that the phenomena related to the quantum Cheshire cat are ``average'' effects. 
In particular, the weak values indicate average shifts of the meter,
conditioned on the pre-selected and the post-selected states, and the object and the property (or two properties of the same object) do not actually travel separately in each arm of the experimental setup; rather they do so only on an average. 
Nonetheless, just like for the original quantum Cheshire cat, in our setups also, it has been ensured that the weak values observed are not classical averages, but quantum-mechanical mean shifts, obtained by considering coupling between the photon or polarization component and a meter. Let us also note here that the amplifications reported  do not imply that the amplified values could become arbitrarily large; nonlinear effects appear that limit the amplified values~\cite{Limits7}.\par 
%
%

\appendix
\section{Effective observable when meter state is continuously distributed}
\label{discrete}
For the initial meter state given in Eq.~(\ref{eq:meter}), the final meter state will take the form
\begin{eqnarray}
  \ket{\Phi_f}_{con}  &\approx&  \braket{\Psi_f|\Psi_{in}}\int dq \, e^{-\frac{q^2}{4\Delta^2}}\Big[1+i g q(\hat{\sigma}_z)_w\nonumber\\ 
  &-& i g^{\prime} t (\hat{L}_x\otimes \hat{\sigma}_x)_w-gg^\prime tq (\hat{L}_x\otimes \hat{\sigma}_x \hat{\sigma}_z)_w \nonumber\\
  &-& g^{\prime^2}\frac{t^2}{2}(\hat{L}_x\otimes \hat{\sigma}_x)_w^2\Big]\ket{q}. 
  \end{eqnarray}
  This can be written as
\begin{equation}
 \ket{\Phi_f}_{con} \approx \braket{\Psi_f|\Psi_{in}}\int \,dq\, e^{iqgA_w}\exp(-\frac{q^2}{4\Delta^2})\ket{q},
  \end{equation}
  with the same $a_w$ and $A^{\prime}_w$ as in Eq.~(\ref{eq:weak_noise}) and hence, the effective observable $A^{\prime}$ is the same as in Eq.~(\ref{eq:A_w_noisy}).
So, the final state of the meter turns out to be
\begin{equation}
 \ket{\Phi_f}_{con} \approx \braket{\Psi_f|\Psi_{in}}\int \,dq\,e^{a_w} e^{iqgA^{\prime}_w}\exp(-\frac{q^2}{4\Delta^2})\ket{q},
\end{equation}
with the corresponding $p$-representation being
\begin{equation}
 \ket{\Phi_f}_{con} \approx \braket{\Psi_f|\Psi_{in}}\int \,dp\,e^{a_w} \exp[-\Delta^2(p-gA^{\prime}_w)^2]\ket{p}.
\end{equation}

\acknowledgements
A.G. and U.S. acknowledge partial support from the Department of Science and Technology, Government of India through QuEST grant (grant number DST/ICPS/QUST/Theme-3/2019/120).

\end{document}